\documentclass[twoside]{article}
\usepackage[latin1]{inputenc}
\usepackage{t1enc}
\usepackage{a4}
\usepackage{tabularx}
\usepackage{epsf}
\usepackage[dvips]{epsfig}

\def\bref{\vspace{4pt}\noindent\hangindent=10mm}

\begin{document}

\setcounter{figure}{0} \setcounter{section}{0}
\setcounter{equation}{0}

\begin{center}
  {\Large\bf $\omega$-Cen -- an Ultra Compact Dwarf Galaxy ?\\[0.7cm]}
  
  Michael Fellhauer\\[0.17cm]
  Inst. Theor. Phys. \& Astrophys. \\
  University of Kiel, Germany \\
  mike@astrophysik.uni-kiel.de
\end{center}

\vspace{0.5cm}

\begin{abstract}
\noindent{\it
  A new formation scenario for $\omega$-Cen is presented, in which
  $\omega$-Cen has formed from a star cluster complex ($=$
  super-cluster).  The star-burst which has led to the formation of 
  this cluster of star clusters was triggered by the infall and
  destruction of a gas-rich dwarf galaxy during the early build-up
  phase of the Milky Way.  While the star clusters in the
  super-cluster merge on a few crossing times scale, which is always
  longer than the individual cluster formation time-scale, they
  capture stars from the destroyed dwarf galaxy very efficiently.
  This process makes it possible to explain the age-spread, the
  spatial and the kinematical differences of the populations observed
  in $\omega$-Cen.  This formation scenario places $\omega$-Cen into
  the same class of objects as the ultra-compact dwarf galaxies (UCDs)
  and the faint fuzzy star clusters (FFs).  }
\end{abstract}

\section{Introduction}
\label{sec:intro}

Hubble Space Telescope images of the central region of the Antennae
colliding galaxies (NGC~4038/4039) (Whitmore et al.\ 1999) reveal
star-bursting regions which have not only formed single young massive
star clusters (YMC or super stellar clusters, SSC) but dozens or even
hundreds of these star clusters together in confined areas (Knots)
forming actually star cluster complexes.  We refer to these complexes
as super-clusters (Kroupa 1998).

One very prominent example is KnotS.  With its age of about $7$~Myr it
is in a stage where most of the star clusters have already merged in
the centre, forming a super stellar cluster spanning several hundred
parsec in radius.  The shape of the surface-brightness profile of
KnotS is exponential in the central region and follows a power law in
the outer parts. 

But these super-clusters are not only found in the centres of strongly
interacting and star-bursting galaxies.  They are also found in 
tidal tails, pulled out from gas-rich interacting galaxies.
Furthermore Larsen et al.\ (2002) found an example of such a
super-cluster in an ordinary, quiet disc galaxy (NGC~6946).  This
super-cluster contains one very massive ($>10^{6}$~M$_{\odot}$) star
cluster and more than a dozen smaller clusters.  All together this
object has a mass in stars of about $10^{7}$~M$_{\odot}$.

These super clusters are already beyond their gas-expulsion phase and
are old enough ($\geq 10$~Myr) to be less centrally concentrated if
they were simply expanding.  Therefore it is very likely that the star
clusters inside these objects are bound to each other.  So far no data
is available of the velocity dispersion of the clusters in the
super-clusters, but the difference of mean radial velocity of two
clusters in the same super-cluster in the Antennae, measured by Mengel
et al.\ (2002), of about $20$~kms$^{-1}$ is not in contradiction with
this assumption.

In a pioneering investigation of the dynamical evolution of such
objects, Kroupa (1998) showed that the constituent clusters in these
star cluster complexes merge in a few super-cluster crossing-times
forming compact spheroidal dwarf galaxies that may have a high
specific frequency of globular clusters and capture field stars
leading to complex chemical- and age-distributions.

At a recent survey of star clusters and dwarf galaxies around the
central galaxy of the Fornax cluster, Hilker et al.\ (1999) and
Phillipps et al.\ (2001) found small, bright (absolute $B$-band
magnitudes of $-13$ to $-11$) and extended objects, too large for
globular clusters and too small for dwarf galaxies.  Shape and
appearance make these objects very small counterparts to M~32.

They call them Ultra-Compact Dwarf galaxies with sizes of about
$100$~pc and effective radii (half-light radii) of $10$-$30$~pc
(Drinkwater et al.\ 2003).  They have velocity dispersions of about
$30$~kms$^{-1}$ implying mass to light ratios of $2$ to $4$, which
leaves no room for dark matter.

In a survey of S0 galaxies Larsen \& Brodie (2000) found intermediate
aged, faint, metal-rich star clusters (Faint Fuzzies) with large
effective radii of greater than $7$~pc rotating within the disc.  One
of the S0 galaxies has a nearby dwarf companion which could have
triggered star formation during a tidal interaction.

In a recent paper Bekki et al.\ (2003) argue that the UCDs could
be the cores of dwarf galaxies which got rid of their envelope and
dark matter halo due to tidal stripping.  According to them UCDs are
the cores of destroyed nucleated dwarf galaxies.  Still they admit
that these cores could have formed out of the merging of several star
clusters.

Furthermore, $\omega$-Cen with its spread in age and metalicity leads
to a new formation scenario with $\omega$-Cen as an UCD (Zinnecker et
al.\ 1988, Tsuchiya et al.\ 2003, Bekki \& Freeman 2003).  According to
these authors $\omega$-Cen may be the core of a nucleated dwarf galaxy
which got stripped of its dark matter halo and the stars in the
envelope, placing the ``naked'' core at the position of $\omega$-Cen
today.  This scenario needs exquisite fine tuning between the orbital
shrinkage rate due to dynamical friction and mass-loss from the
satellite such that only the core is left without dark matter and
stellar envelope. 

$\omega$-Cen is the most massive globular cluster in the Milky Way.
It contains about $5 \cdot 10^{6}$~M$_{\odot}$ (Meylan et al.\ 1995)
and it orbits the Milky Way in a slightly inclined (mostly within the
disc), retrograde orbit well within the Solar radius.  It shows
rotation with a maximum rotation speed of $8$~kms$^{-1}$ (Freeman
2001) and is slightly flattened.  Some observational properties of
$\omega$-Cen can be found in Tab.~\ref{tab:ocen}. 
\begin{table}[h!]
  \begin{center}
    \begin{tabular}[h!]{ll} \hline
      galactocentric distance   & $6.7$~kpc \\
      total luminosity          & $M_{V}=-10.3$~Mag. \\
      total mass ($M/L=4.1$)    & $5.1 \cdot 10^{6}$~M$_{\odot}$
      \\ 
      central surface brightness & $\mu_{V0}= 16.77$ mag./arcsec$^{2}$
      \\
      core radius               & $3.7$~pc \\
      half-mass radius          & $6.1$~pc \\
      tidal radius              & $64.6$~pc \\
      flattening                & $\epsilon = 0.121$ \\
      velocity dispersion       & $21.9$~km/s \\
      maximum rotation velocity & $8$~km/s \\
      metalicity                & -1.68 (mean); -1.8 to -0.8 \\
      age                       & 15 Gyr; age spread $\approx$ 4 Gyr
      \\
      \hline
    \end{tabular}
    \caption{Observational Properties of $\omega$-Cen (taken from
      various authors).}
    \label{tab:ocen}
  \end{center}
\end{table}

The most interesting property of $\omega$-Cen is its metalicity and
age distribution.  It has a mean metalicity of $[$Fe/H$] = -1.68$
(Meylan et al. 1995) and exhibits three different metalicity peaks
(Hilker \& Richtler 2000).  But the stars do not only show a spread in
metalicity but also in age.  The estimated values for the age-spread
vary in the range of $3$--$5$~Gyr. 

Finally the different populations of $\omega$-Cen show differences in
their spatial distribution and kinematics.  The older metal-poor stars
are rotating while the younger metal-rich stars are more centrally
concentrated and show no signs of rotation.

While the study reported here agrees with the suggestion that
$\omega$-Cen is an UCD (Fellhauer \& Kroupa 2002a,b), a different
scenario for the formation is proposed.  If a gas-rich dwarf
galaxy fell into the Milky Way during the early build-up phase,
this infall could have triggered a local star-burst within the
dissolving dwarf forming a star cluster complex ($=$ super-cluster).
The star clusters in this super-cluster then merge and form a compact
merger-object.  The large age-spread and the spread in metalicities
can be accounted for by the capture of an underlying population of old
stars, which surround the super-cluster and have the same orbit.
These stars can be leftovers of the destroyed dwarf galaxy which
triggered the star burst.  They can not be stars from the Milky Way
disc because of the retrograde orbit of $\omega$-Cen.  The relative
velocity between the Milky Way stars and $\omega$-Cen is too large to
capture them.  But stars from the destroyed dwarf galaxy, having the
same overall streaming motion, could be easily captured by the
potential well of the newly forming merger-object.  The main
difference between this model and the ``standard'' infall model of
$\omega$-Cen as the stripped core of a dwarf galaxy is that instead of
having formed as the core of a dwarf galaxy before the infall, it is
formed when the dwarf galaxy falls into the Milky Way.  $\omega$-Cen
would be born off-centre, with the surrounding dwarf galaxy in the
stage of dissolution.  Thus there is no need for the degree of
fine-tuning required by the ``standard'' infall model.

In Section~\ref{sec:setup} the setup of the models is explained
followed by a brief description of the simulation code {\sc Superbox}
(Fellhauer et al.\ 2000) in Section~\ref{sec:sbox}.
Section~\ref{sec:res} then gives the results of the previous work,
modelling UCDs and FFs followed by the best fit model of $\omega$-Cen
so far.  Finally I conclude with a comparison of the results achieved
so far with the ``standard'' infall-model of $\omega$-Cen.

\section{Setup of the Simulations}
\label{sec:setup}

A super-cluster is calculated orbiting in an analytical galactic
potential (disc + halo).  The halo is modelled as a logarithmic
potential and the disc potential has the form of a Plummer-Kuzmin
disc:
\begin{eqnarray}
  \label{eq:galpot}
  \Phi_{\rm gal} & = & \Phi_{\rm disc} + \Phi_{\rm halo}
   \nonumber \\
   & = & - \frac{GM_{\rm disc}}{\sqrt{R^{2} + ( a +
      \sqrt{Z^{2}+b^{2}})^{2} } } - \frac{1}{2} v_{0}^{2} \ln(
  R_{\rm gal}^{2} + R^{2})
\end{eqnarray}
with $M_{\rm disc} = 10^{11}$~M$_{\odot}$, $a = 3$~kpc, $b = 0.3$~kpc,
$v_{0} = 200$~km/s and $R_{\rm gal} = 50$~kpc which sums up to an
almost flat rotation curve with a rotation speed of $220$~kms$^{-1}$.

The super clusters are Plummer-distributions with characteristic radii
$r_{\rm pl}^{\rm sc}$ ranging from $20$ to $500$~pc, a cut-off radius of
$r_{\rm cut}^{\rm sc} = 5 r_{\rm pl}^{\rm sc}$ and with masses $M_{\rm
  sc} = 10^{4}$ to $2 \cdot 10^{7}$~M$_{\odot}$ containing 20 to 262
star clusters of equal masses or following a power law distribution
\begin{eqnarray}
  \label{eq:massdist}
   n(M_{\rm cl}) & \propto & M_{\rm cl}^{-1.5},
\end{eqnarray}
which mimics the mass-spectrum for young star clusters found in most
star bursting galaxies (e.g.\ Zhang et al.\ 1999).

Each star cluster is itself modelled as a Plummer-sphere with $10000$
to $100000$ particles and a Plummer-radius (half-light radius) of
$r_{\rm pl} = 4$~pc, which is the mean effective radius found in the
young star cluster population in the central part of the Antennae
(Whitmore et al.\ 1999).

For the underlying old population, used for the $\omega$-Cen models,
a homogeneous ellipsoid of stars, with a density below the background
density ($=$ density of the Milky Way) is modelled.  The stars have
velocities according to the epicyclic approximation, which leaves them
unbound to each other but forming a quasi-stable configuration in
phase-space.  This ellipsoid of stars mimics a region of stars from
the destroyed dwarf galaxy. 

\section{ The {\sc Superbox}-Code}
\label{sec:sbox}

The simulations are performed with the particle-mesh code {\sc
  Superbox} (Fellhauer et al.\ 2000). In a particle-mesh technique
particles are assigned to densities on a Cartesian grid. These
densities are then converted to a potential via the Fast Fourier
Transform. Forces are calculated from the potential using a
higher-order nearest grid-point scheme. The particle velocities and
positions are then integrated using a fixed-time-step leapfrog scheme.

The feature, which makes {\sc Superbox} suitable for these kinds of
simulations are the two levels of high-resolution sub-grids tracing
each object throughout the simulation volume providing high-resolution
only at the places where it is necessary.  The lowest grid resolution
covers the whole simulation volume with the orbit around the host
galaxy.  The medium level grids cover the volume of the super-cluster
resolving the forces between the star clusters.  Finally the highest
resolution grids cover the individual star clusters and resolve the
internal forces and the forces in the merging of two clusters.

\section{General Results}
\label{sec:res}

To examine how fast and how effective the star clusters merge, a set
of simulations with super-clusters with different richness,
\begin{eqnarray}
  \label{eq:alpha}
  \alpha & = & \frac{r_{\rm pl}}{r_{\rm pl}^{\rm sc}} \ ,
\end{eqnarray}
(varying not only the size of the super-cluster but also the size of
the star clusters) and different distances from the centre of their
host galaxy mimicking different strengths of the tidal field,
\begin{eqnarray}
  \label{eq:beta}
  \beta  & = & \frac{r_{\rm pl}^{\rm sc}}{r_{\rm tidal}} \ ,
\end{eqnarray}
are performed.  $r_{\rm tidal}$ denotes the tidal radius of the
super-cluster (Binney \& Tremaine 1987),
\begin{eqnarray}
  \label{eq:rtidal}
  r_{\rm tidal}(D) & = & \left( \frac{M_{\rm sc}} {3M_{\rm
        gal}(D)} \right)^{1/3} D,
\end{eqnarray}
where $D$ denotes the distance to the centre of the galaxy and $M_{\rm
  gal}(D)$ is the enclosed mass of the galaxy at that distance.

\begin{figure}[t!]
  \begin{center}
    \epsfxsize=05.8cm \epsfysize=05.8cm \epsffile{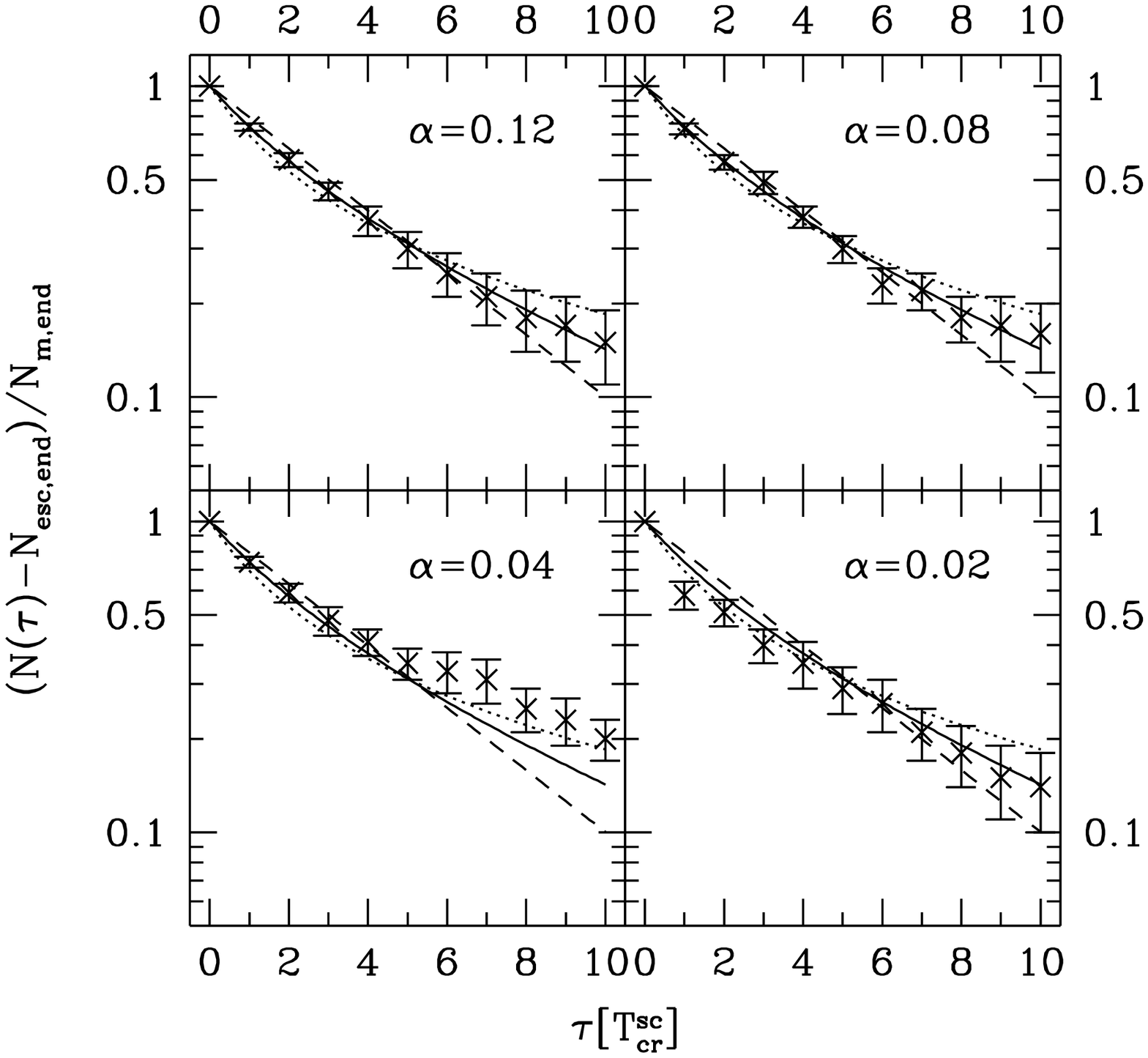}
    \epsfxsize=05.8cm \epsfysize=05.8cm \epsffile{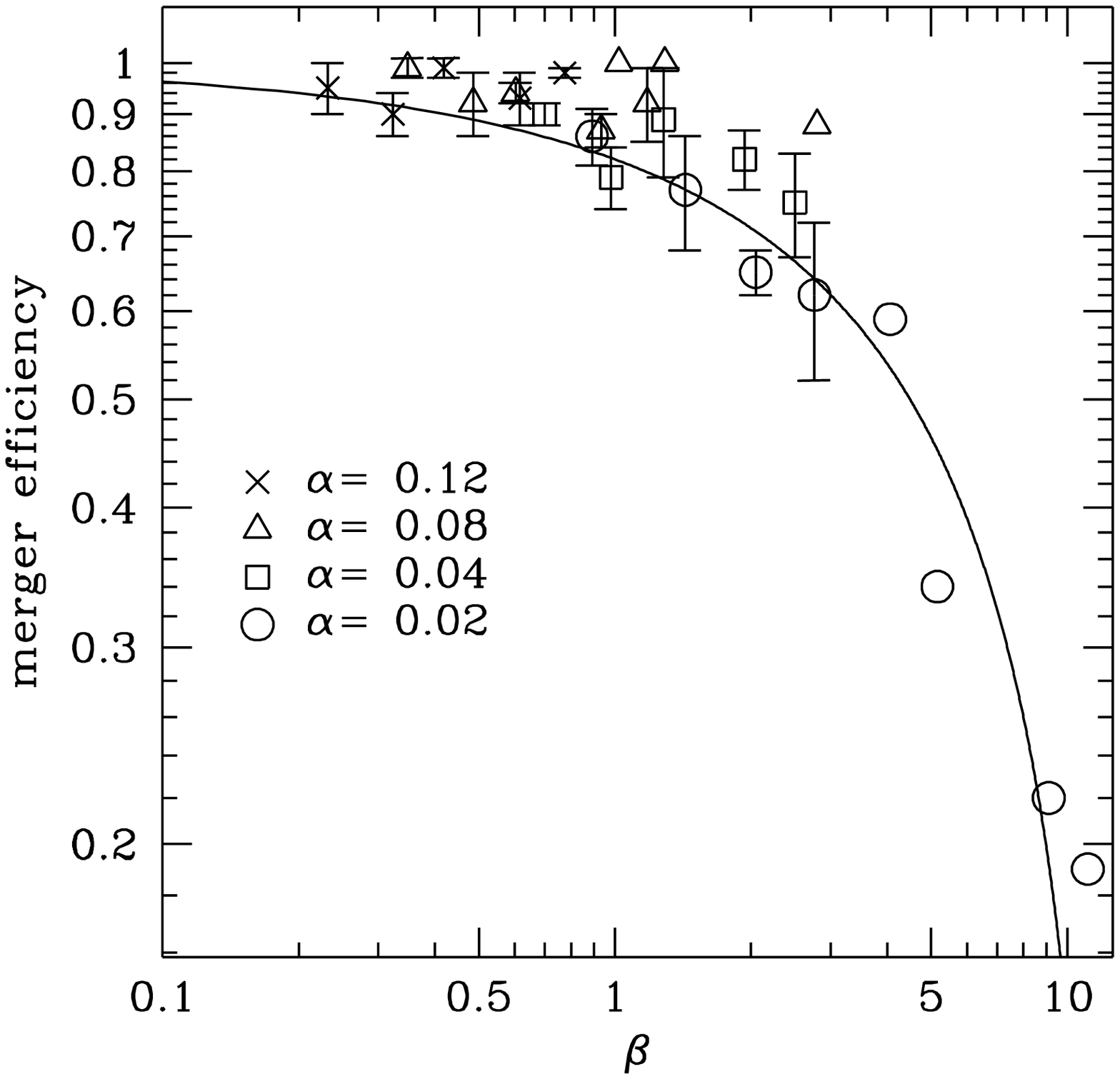}
    \caption{Left: Merging time-scales for the different
      richnesses of the super-clusters $\alpha$.  Note that the
      fitting curve (Eq.~\ref{eq:mtimesol}) is independent of the
      richness of the cluster.  Right: Merger-efficiency as function
      of the strength of the tidal field $\beta$.  The data shows that
      the tidal field has to be very strong to inhibit star clusters
      from merging.}
    \label{fig:timeeff}
  \end{center}
\end{figure}

The merging behaviour is best described as two concurrent processes.
The major process is the merging between the star clusters and the
merger object and the second process is the merging of two star
clusters.  Following the arguments of Fellhauer (2001) one gets the
following differential equation for the evolution of the number of
star clusters $N$:
\begin{eqnarray}
  \label{eq:mtime}
  -\frac{{\rm d} N}{{\rm d}\tau} & = & d_{1} N^{2} + d_{2} N
\end{eqnarray}
The time $\tau$ is measured in internal crossing-times, $T_{\rm
  cr}^{\rm sc}$, of the super-cluster, the first term on the right-hand
side corresponds to the merging of two clusters ($\propto N^{2}$), the
second to the major process - the merging with the merger object
($\propto N$).  The solution to Eq.~\ref{eq:mtime} is
\begin{eqnarray}
  \label{eq:mtimesol}
  \frac{N(\tau)} {N_{0}} & = & \frac{\exp(-d_{2}\tau)} {1 + \chi (1 - 
    \exp(-d_{2}\tau))}
\end{eqnarray}
with $\chi = d_{1}N_{0}/d_{2}$, and $N_{0}$ denotes the number of star
clusters at $\tau = 0$.  Fitting this formula to the data gives the
following values for $d_{2}$ and $\chi$ (see also left panel of
Fig.~\ref{fig:timeeff})
\begin{eqnarray}
  \label{eq:solu}
  d_{2} & = & 0.11 \pm 0.01 \ ,\\
  \chi  & = & 2.0  \pm 0.3 \ . \nonumber
\end{eqnarray}

The merger efficiency can be described with the following fitting
formula (see also right panel of Fig.~\ref{fig:timeeff}):
\begin{eqnarray}
  \label{eq:mrate}
  \frac{N_{\rm m}}{N_{0}} & = & 1 - 0.18 \cdot \beta^{0.68}\ ,
\end{eqnarray}
which shows only very strong tidal fields ($\beta \gg 1$) could
inhibit the star clusters from merging with each other.

\begin{figure}[t!]
  \begin{center}
    \epsfxsize=05.8cm \epsfysize=05.8cm \epsffile{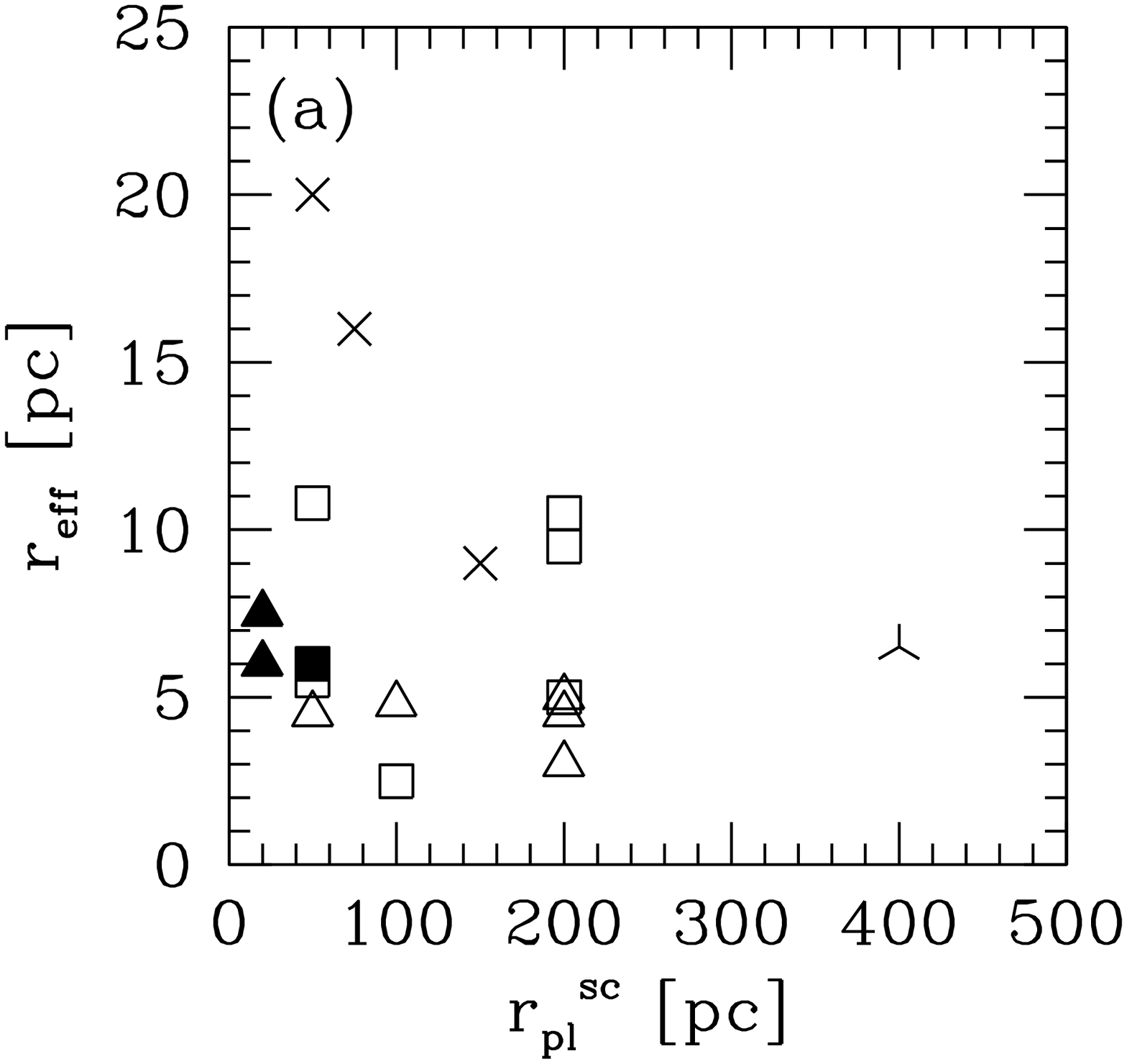}
    \epsfxsize=05.8cm \epsfysize=05.8cm \epsffile{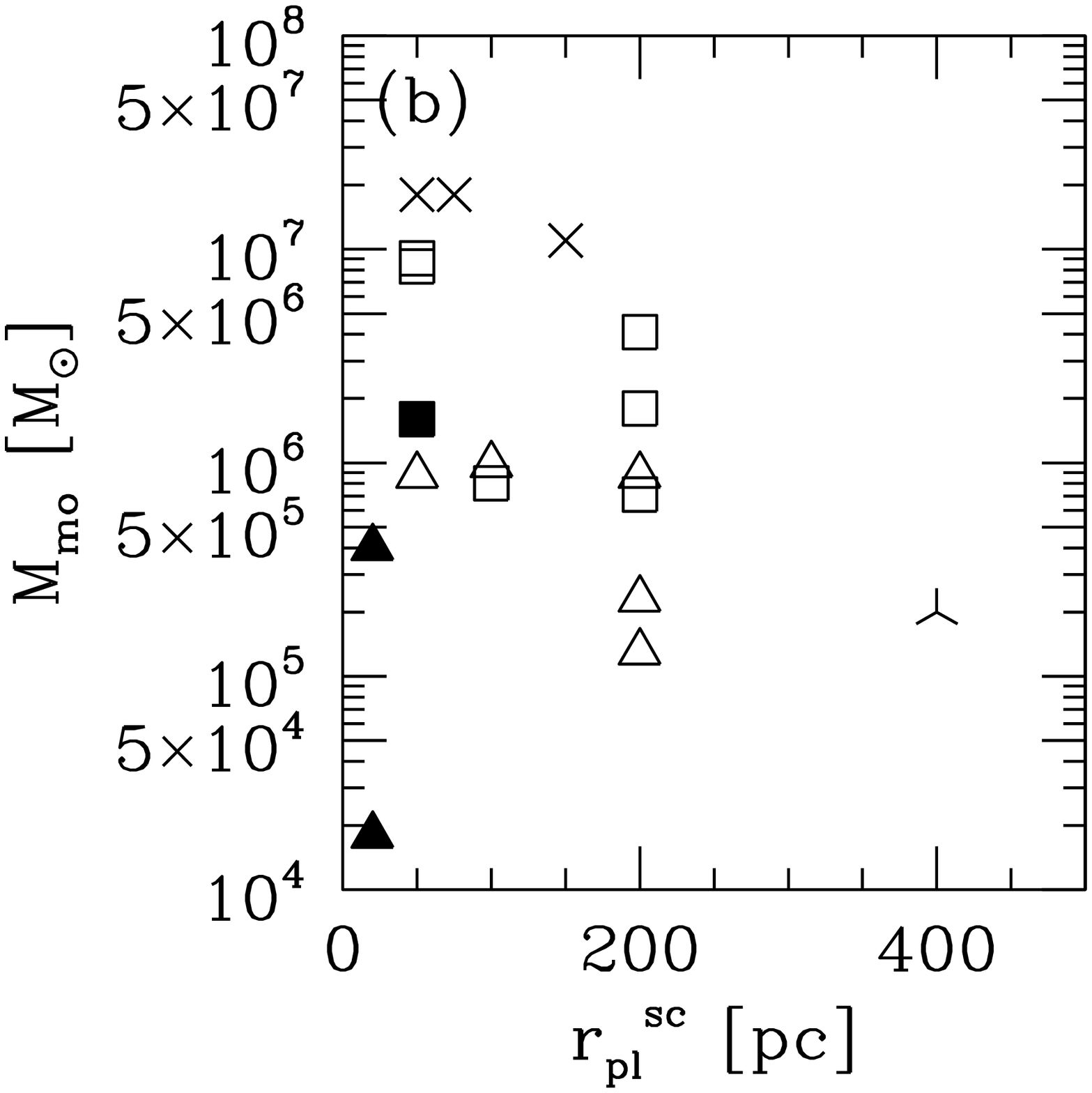}
    \caption{Left: Effective radius of the merger-object as a function
      of the scale-length of the super-cluster.  Right: Mass of the
      merger-object as a function of the scale-length of the
      super-cluster.  For both panels: Crosses are massive
      super-clusters in a weak tidal field whose merger objects have
      large effective radii and are massive (like UCDs).  Boxes are
      massive super-clusters in strong tidal fields.  Their
      merger-objects have smaller effective radii but are still
      massive.  Triangles are low-mass super-clusters in a strong
      tidal field.  Their merger objects are similar to the FF.} 
    \label{fig:eradmass}
  \end{center}
\end{figure}

An extension to eccentric orbits following the evolution
of the merger-objects over several Gyr yields virtually the same
results.  All the merger-objects are compact but have extended
effective radii, which make the smaller ones differ from ordinary
globular clusters and the larger ones look similar to the newly
discovered UCDs.  An overview of masses and effective radii is given
in Fig.~\ref{fig:eradmass}.  The objects are very stable against tidal
disruption and survive for a Hubble-time. 

The sizes of the merger objects range from several dozens to several
hundreds of parsec depending on the strength of the tidal field and
how effective the merger process was.  In the case of strong tidal
forces and extended super-clusters the merger efficiency is relatively
low.  Only a few clusters merge and build one small or sometimes two
small merger objects.  These objects have sizes and masses comparable
to globular clusters (GC).  This process could act as an additional
destruction process for low mass clusters.  It adds high mass clusters
to the present day globular cluster mass function.  

Simulations are performed to resemble Faint Fuzzy star clusters out of
this merging scenario.  The super-clusters have to be inside the disc,
are not very rich and massive and experience a strong tidal field.
The resulting merger objects have masses of $10^{4}$ to
$10^{6}$~M$_{\odot}$ and show large effective radii of $> 6$~pc.  The
surface density profile can be best fitted with an exponential in the
inner part and a steep power-law profile in the outer part or with a
King profile with a large core radius ($=$ effective radius).  A
detailed discussion of the results can be found in Fellhauer \& Kroupa
(2002b). 

Placing rich, massive super-clusters in the outer part of the
host-galaxy (halo), where they experience only a weak tidal field, the
models resemble objects which look quite similar to the observed ultra
compact dwarf galaxies.  The merger-objects have sizes of $> 100$~pc
and masses $> 10^{6}$~M$_{\odot}$.  They show effective radii of $>
10$~pc and velocity dispersions of $20$ -- $30$~kms$^{-1}$.  The
surface density profile could again be fitted with a King profile or
with an exponential in the innermost part, a power-law proportional to
$r^{-2}$ in the medium range out to the tidal radius at perigalacticon
and a much steeper power law ($\propto r^{-4.5}$) beyond this point.
This can be understood by this material getting unbound near
perigalacticon.  But the stars do not leave this region immediately
and some are recaptured, when the object is near apogalacticon.  These
objects are discussed in detail in Fellhauer \& Kroupa (2002a).

\section{$\omega$-Cen}
\label{sec:ocenres}

The setup of this $\omega$-Cen model is realized with a super-cluster
with a Plummer radius $r_{\rm pl}^{\rm sc} = 20$~pc, with a total mass
of $10^{7}$~M$_{\odot}$ containing $20$ massive young star clusters
modelled as Plummer-spheres with $100000$ particles each.  In addition,
an underlying old population with $500000$ particles in a
homogeneous ellipsoid is modelled with $1/10$ of the background
density of the host-galaxy, having a small half-axis of $500$~pc.  The
stars of this population are not bound to each other and have orbits
according to the epicyclic approximation.  For simplicity the
super-cluster is placed on a circular orbit around the host-galaxy at
a distance of $5$~kpc.

The star clusters of the super-cluster merge completely within
$30$~Myr which corresponds to $\approx 12$ crossing times of the
super-cluster.  In the same time almost all of the unbound stars are
captured by the potential well of the merger-object.

\begin{figure}[t!]
  \begin{center}
    \epsfxsize=08.0cm \epsfysize=08.0cm \epsffile{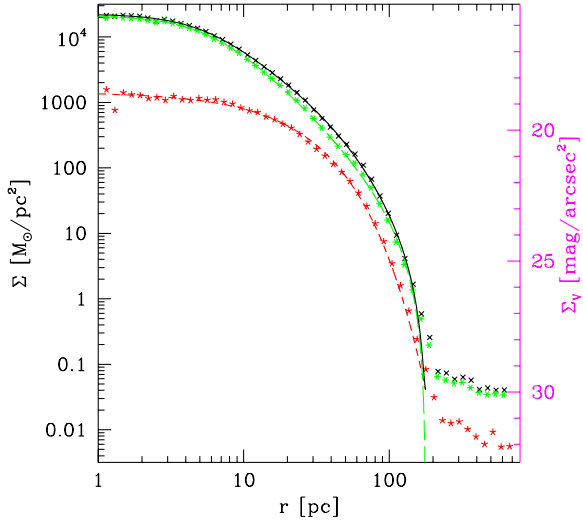}
    \caption{Surface density profile of the merger-object.  Crosses
      show the profile of all stars (fitted with a King profile),
      five-pointed stars are the older stars from the disrupted dwarf
      galaxy (exponential fit) and six-pointed stars are the younger
      stars from the super-cluster (again King profile).  Parameters
      of the  fitting curves are given in the main text.  Right
      abscissa denotes surface-brightnesses using the mass-to-light
      ratio of $\omega$-Cen.}
    \label{fig:surf}
  \end{center}
\end{figure}

\begin{figure}[t!]
  \begin{center}
    \epsfxsize=05.8cm \epsfysize=05.8cm \epsffile{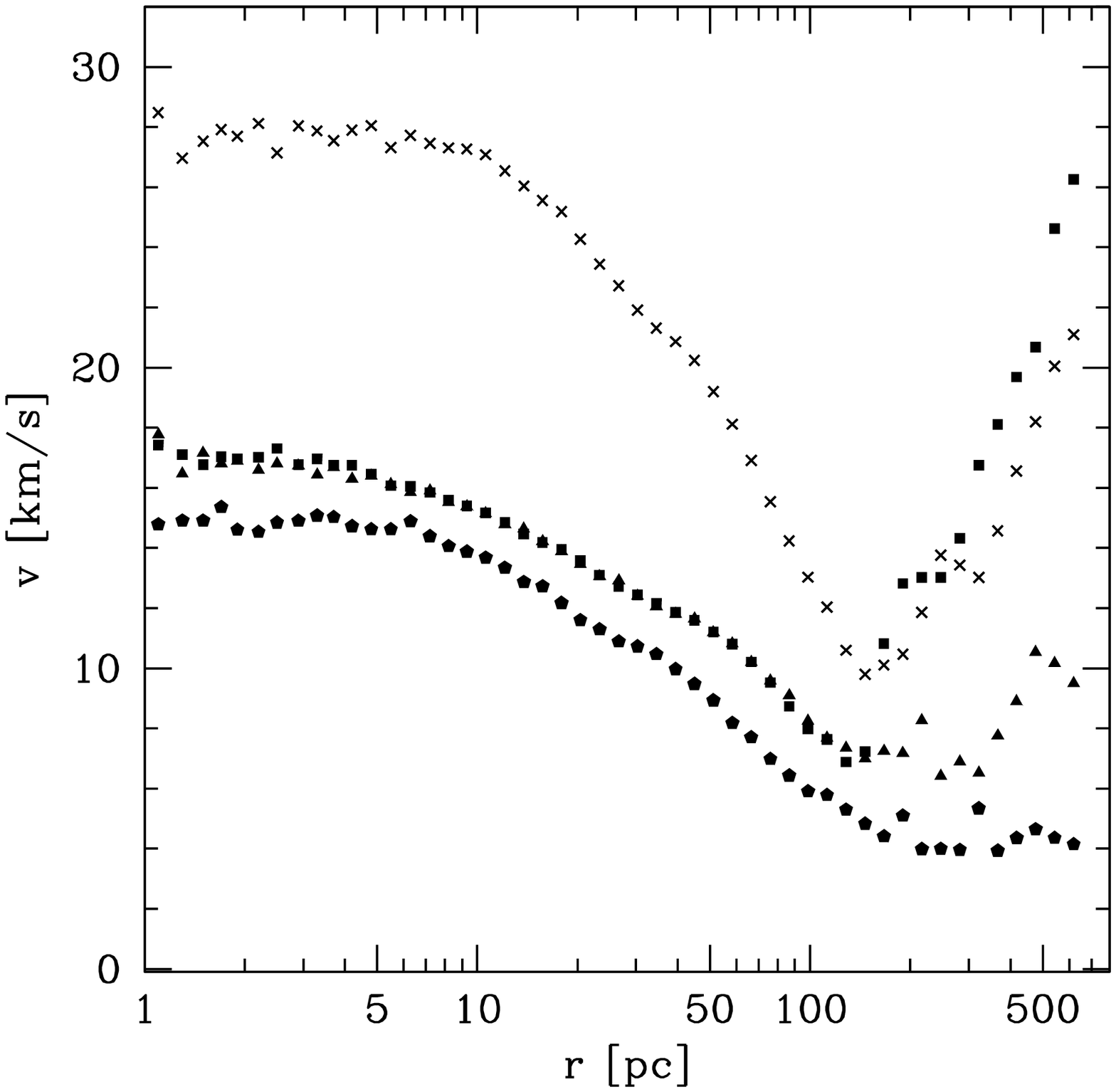}
    \epsfxsize=05.8cm \epsfysize=05.8cm \epsffile{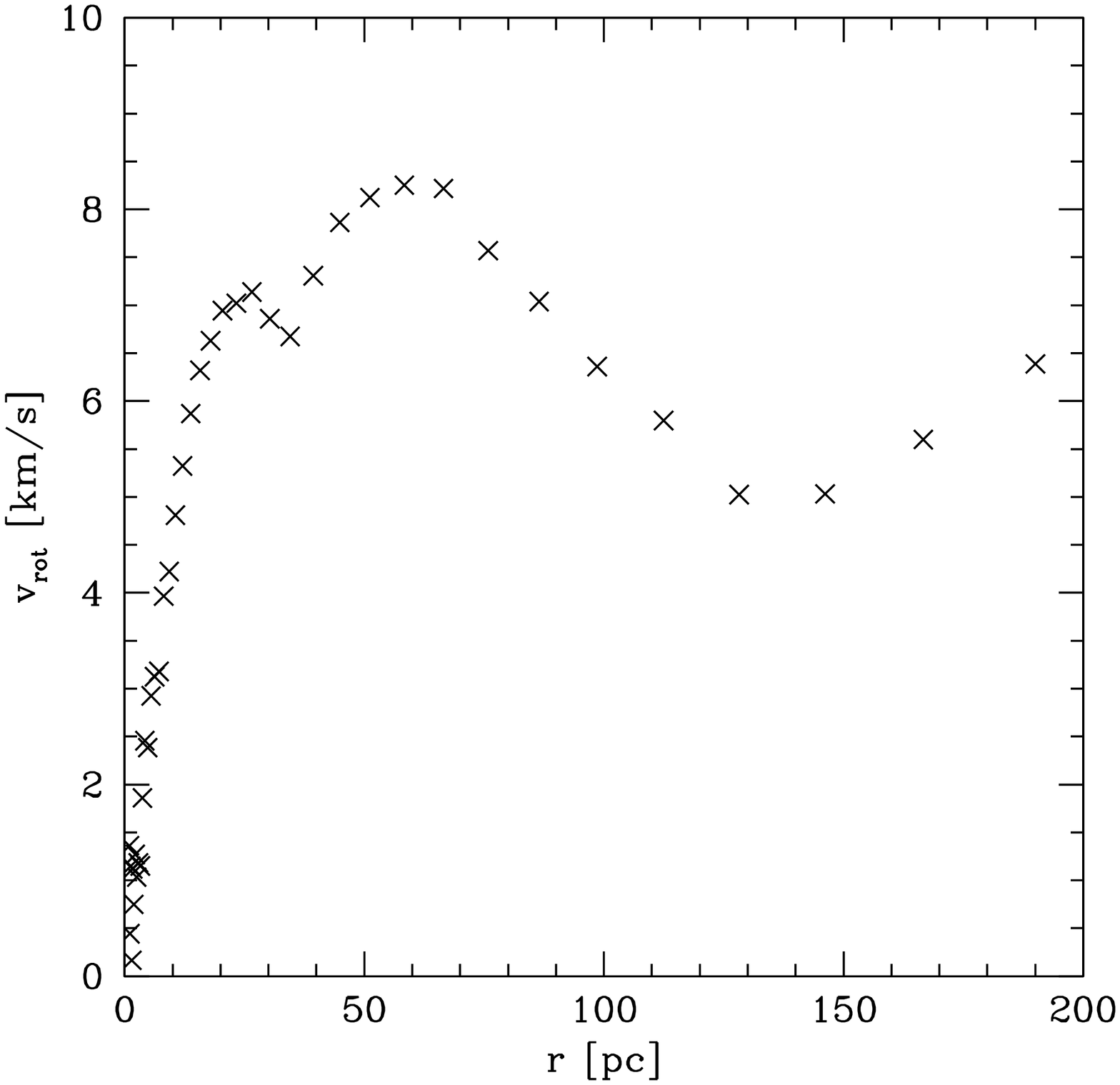}
    \caption{Left: Velocity dispersion profile.  From top to bottom:
      3-D velocity dispersion (measured in concentric spherical
      shells), $x$-, $y$- and $z$-direction of the line-of-sight
      velocity dispersion (measured in concentric cylinders). Right:
      Rotation curve of the merger object.}
    \label{fig:velrot}
  \end{center}
\end{figure}

Because of the circular orbit this merger-object does not loose mass
efficiently and after $5$~Gyr the merger-object still
has $8 \cdot 10^{6}$~M$_{\odot}$ or, assuming the mass-to-light ratio
of $4.1$ of $\omega$-Cen, a total visual brightness of
$M_{V}=-10.86$~Mag.  This is much too massive compared to
$\omega$-Cen and will be changed in future models with more realistic
orbits for the merger-objects.  Therefore the size of the
merger-object also does not correspond to the size of the real
$\omega$-Cen.  In the following the shape of the profiles are
discussed rather than the actual extension.  The present model thus
represents a first try with improvements to follow (Fellhauer \&
Kroupa, in prep.).

\begin{figure}[t!]
  \begin{center}
    \epsfxsize=08.0cm \epsfysize=08.0cm \epsffile{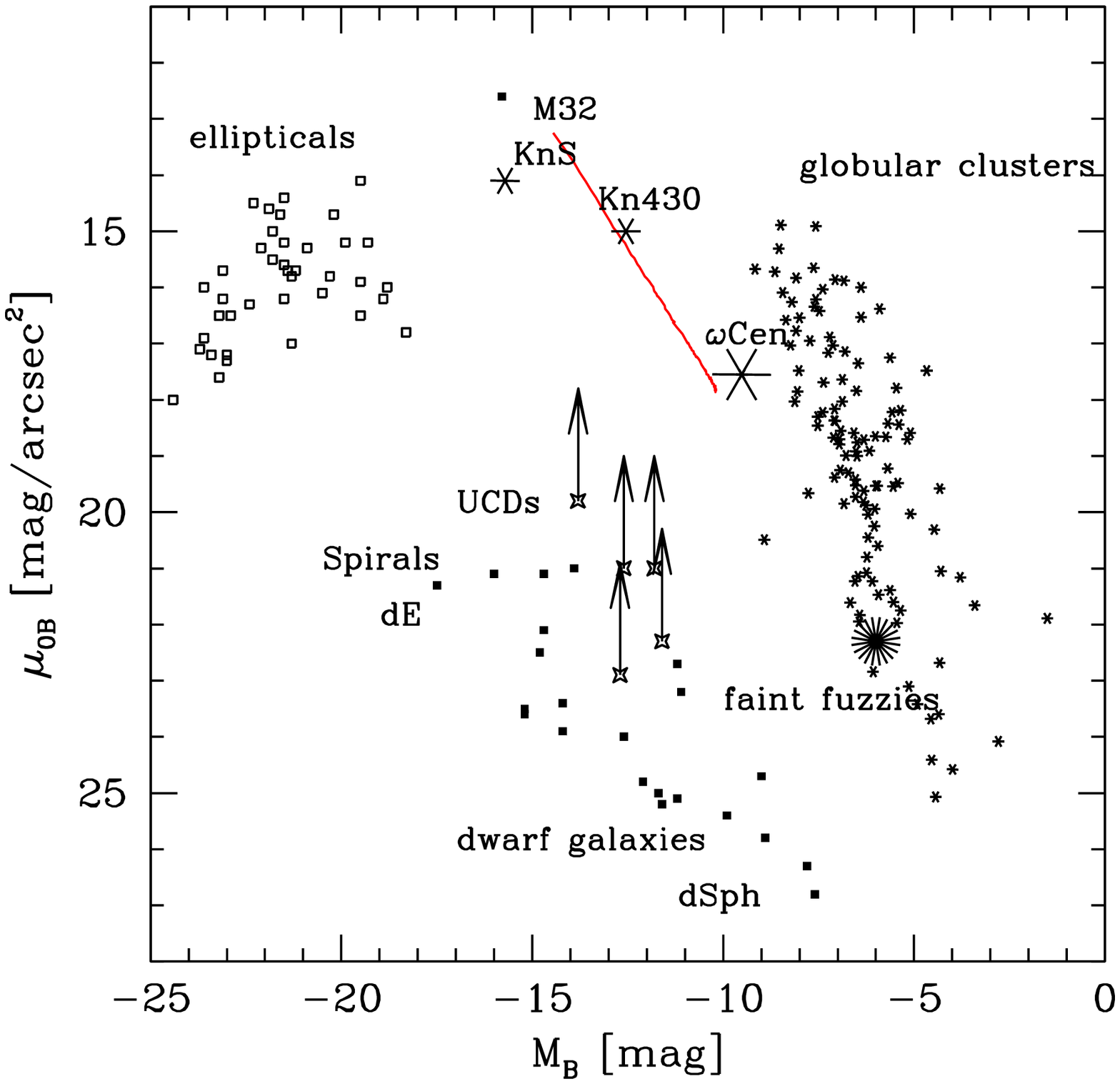}
    \caption{Kormendy diagram with the evolution of the merger
      object until 5 Gyr.  Central surface brightness, $\mu_{0,{\rm
          B}}$, vs absolute photometric B-band magnitude, $M_{\rm B}$
      (the Kormendy diagram).  The stars are data for Milky-Way
      globular clusters (GC, Harris 1996), Local-Group dwarf galaxies
      (Mateo 1998) are displayed as filled boxes, elliptical galaxies
      (E, Peletier et al.\ 1990) are shown as open boxes.  The
      positions of disk galaxies are indicated by ``Spirals'' (cf.
      Ferguson \& Binggeli 1994).  The newly discovered UCDs are shown
      by arrows with lower-limits on $\mu_{\rm B}$ (Phillipps et al.\ 
      2001). Two of the most massive ``knots'', (KnS, Kn430, tables~1
      and~2 in Whitmore et al.\ 1999), observed in the interacting
      Antennae galaxies are shown as stars.  Knots KnS and Kn430 are
      roughly $10$~Myr old so that $M/L_{\rm B} \approx 0.035$.  The
      large fuzzy symbol shows the approximate location of the faint
      fuzzy star clusters.  The solid line marks the evolution of the
      $\omega$-Cen model discussed in the text.}
    \label{fig:korm}
  \end{center}
\end{figure}

Fig.~\ref{fig:surf} shows the surface density profile of the
merger-object after $5$~Gyr of evolution.  The solid line shows the
profile of all particles, which can be best fitted with a King-profile
with a central surface density of about $24000$~M$_{\odot}$pc$^{-2}$ and
a core radius of about $6$~pc.  This corresponds to a central surface
brightness of $17$~mag.arcsec$^{-2}$.  As mentioned above, due to the fact
that the merger-object is too massive the tidal radius of $170$~pc is
too large.  The long dashed line shows the profile of the younger
population from the merged star clusters.  Their best fit is again a
King profile with a core radius of $5.8$~pc.  The central properties
of the merger-object differ somewhat from the real $\omega$-Cen,
mainly because the Code does not allow for two-body relaxation.
Therefore the merger-object does not show signs of core-contraction
which would be expected even for a massive object like $\omega$-Cen
with a half-mass relaxation-time close to a Hubble-time.  The older
stars from the underlying population are displayed with the short dashed
line.  They show an extended profile which is best fitted with an
exponential profile with a scale-length of $17$~pc.  While the
absolute numbers do not exactly match $\omega$-Cen, the model shows
qualitatively similar properties.  Like in $\omega$-Cen the older
stars are less centrally concentrated.  Also the shape of the profile
is in good agreement with the latest data from $\omega$-Cen published
by Leeuwin et al.\ (2002). 

Fig.~\ref{fig:velrot} (left panel) shows the velocity dispersion
profile of the model.  It has a 3D-dispersion of about $27$~kms$^{-1}$
and a line-of-sight velocity dispersion of about $17$~kms$^{-1}$ along
the major axis and $15$~kms$^{-1}$ along the minor axis.  This is in
agreement with the data from Leeuwin et al.\ (2002).

Also the model shows rotation in the old population with a maximum
rotation speed of about $8$~km/s (Fig.~\ref{fig:velrot} right panel).
The flattening ($0.06$) is less strong than in $\omega$-Cen ($0.12$).
Compared with the data from Freeman (2001), both the maximum rotation
velocity and the shape of the rotation curve are reproduced.

Finally converting the total mass and central density
into total brightness and central surface brightness using
mass-to-light ratios derived from a single stellar population computed
with Starburst99 (Leitherer et al.\ 1999), and changing to linear
fading at the point where the Starburst99 code becomes unreliable (as
described in Fellhauer \& Kroupa 2001a), the evolution of the
merger-object can be placed into a Kormendy diagram.
Fig.~\ref{fig:korm} nicely shows that the model evolves from the most
massive super-clusters in the Antennae to the approximate location of
$\omega$-Cen today.

\section{Conclusions}
\label{sec:conclus}

A new formation scenario for $\omega$-Cen is presented which differs
from the ``standard'' infall-model where $\omega$-Cen is the core of a
stripped dwarf galaxy.  These ``standard'' models have problems in
finding the right amount of stripping of the dwarf galaxy halo and
envelope to place the naked core at the present orbit of $\omega$-Cen
today. If the envelope gets stripped too fast dynamical friction is
not powerful enough to bring the core to its present position.  In
contrast, the models described here propose that $\omega$-Cen has
formed when the dwarf galaxy fell into the Milky Way and got
destroyed.  The infall caused a star-burst forming star clusters in a
super-cluster travelling on the same orbit as the old field stars of
the destroyed dwarf galaxy.  During the merger-process the
merger-object can capture these old stars quite effectively and the
two different populations in the models have the same shape and are
qualitatively correct when compared with the real data for
$\omega$-Cen.  The old population is less centrally concentrated and
shows the right amount of rotation. The young population is more
concentrated to the centre. 

Anyway which scenario one might prefer, they both have the conclusion
that $\omega$-Cen is indeed an Ultra-Compact Dwarf Galaxy.

\subsection*{References}

{\small
  
  \bref Bekki, K., Couch, W.J., Drinkwater, M.J., Shioya, Y., 2003,
  MNRAS, {\bf 344}, 399.
  
  \bref Bekki, K., Freeman, K.C., 2003, MNRAS accepted,
  astro-ph/0310348.
  \bref Binney, J., Tremaine, S., 1987, ``Galactic Dynamics'',
  Princeton University Press, ISBN 0-691-08445-9.
  
  \bref Brodie, J.P., Larsen, S.S., 2002, AJ, {\bf 124}, 1410.
  
  \bref Bruzual, G.A., Charlot, S., 1993, ApJ, {\bf 405}, 538.
  
  \bref Drinkwater, M.J., Gregg, M.D., Hilker, M., Bekki, K., Couch,
  W.J., Ferguson, J.B., Jones, J.B., Phillipps, S., 2003, Nature in
  press 

  \bref Fellhauer, M., Kroupa, P., Baumgardt, H., Bien, R., Boily,
  C.M., Spurzem, R., Wassmer, N., 2000, NewA, {\bf 5}, 305.
  
  \bref Fellhauer, M., 2001, PhD-thesis, University of Heidelberg,
  Shaker-Verlag, ISBN~3-8265-8658-1.
  
  \bref Fellhauer, M., Kroupa, P., 2002a, MNRAS, {\bf 330}, 642.
  
  \bref Fellhauer, M., Kroupa, P., 2002b, AJ, {\bf 124}, 2006.
  
  \bref Ferguson, H.C., Binggeli, B., 1994, A\&AR, {\bf 6}, 67.
  
  \bref Freeman, K.C., 2001, in proceedings of Star2000: The dynamics
  of star clusters and the Milky Way, ed.\ S. Deiters, B. Fuchs, A.
  Just, R. Spurzem, R. Wielen, San Francisco, California: ASP, ASP
  Conf.\ Ser.\ {\bf 228}, 43.
  
  \bref Harris, W.E., 1996, AJ, {\bf 112}, 1487.
  
  \bref Hilker, M., Infante, L., Kissler-Patig, M., Richtler, T.,
  1999, A\&AS, {\bf 134}, 75.
  
  \bref Hilker, M., Richtler, T., 2000, A\&A, {\bf 362}, 895.
  
  \bref Kroupa, P., 1998, MNRAS, {\bf 300}, 200.
  
  \bref Larsen, S.S., Brodie, J.P., 2000, AJ, {\bf 120}, 2938.
  
  \bref Larsen, S.S., Efremov, Y.N., Elmegreen, B.G., Alfaro, E.J.,
  Battinelli, P., Hodge, P.W., Richtler, T., 2002, ApJ, {\bf 567},
  896.
  
  \bref van Leeuwin, F., Le Pode, R.S., Reijns, R.A., Freeman, K.C.,
  de Zeeuw, P.T., 2000, A\&A, {\bf 360}, 472.
  
  \bref Leitherer, C., Schaerer, D., Goldader, J.D., Delgado, R.M.G.,
  Robert, C., Kune, D.F., de Mello, D.F., Devost, D., Heckman, T.M.,
  1999, ApJS, {\bf 123}, 3.
  
  \bref Mengel, S., Lehnert, M.D., Thatte, N., Genzel R., 2002, A\&A,
  {\bf 383}, 137.
  
  \bref Mateo, M., 1998, ARAA, {\bf 36}, 435.
  
  \bref Meylan, G., Mayor, M., Duquenney, A., Dubath, P., 1995, A\&A,
  {\bf 303}, 761.
  
  \bref Pelletier, R.F., Davies, R.L., Illingworth, G.D., Davies,
  L.E., Cawson, M., 1990, AJ, {\bf 100}, 1091.
  
  \bref Phillipps, S., Drinkwater, M.J., Gregg, M.D., Jones, J.B.,
  2001, ApJ, {\bf 560}, 201.
  
  \bref Tsuchuiya, T., Dinescu, T.I., Korchagin, V.I., 2003, ApJL,
  {\bf 589}, L29.
  
  \bref Whitmore, B.C., Zhang, Q., Leitherer, C., Fall, S.M., 1999,
  AJ, {\bf 118}, 1551.
  
  \bref Zhang, Q., Fall, S.M., 1999, ApJL, {\bf 527}, 81L.  

  \bref Zinnecker, H., Keable, C.J., Dunlop, J.S., Cannon, R.D.,
  Griffiths, W.K., 1988, in proceedings of the 126th Symposium of the
  International Astronomical Union: The Harlow Shapley Symposium on
  Globular Cluster Systems in Galaxies, ed.\ Grindlay, J.E., Philip,
  A.G.D., Dordrecht, NL, Kluwer Academic Publishers, p. 603

  }

\vfill

\end{document}